\documentstyle[12pt,epsf,epsfig]{article}
\newcount\timecount
\newcount\hours \newcount\minutes  \newcount\temp \newcount\pmhours

\hours = \time
\divide\hours by 60
\temp = \hours
\multiply\temp by 60
\minutes = \time
\advance\minutes by -\temp
\def\hour{\the\hours}
\def\minute{\ifnum\minutes<10 0\the\minutes
            \else\the\minutes\fi}
\def\clock{
\ifnum\hours=0 12:\minute\ AM
\else\ifnum\hours<12 \hour:\minute\ AM
      \else\ifnum\hours=12 12:\minute\ PM
            \else\ifnum\hours>12
                 \pmhours=\hours
                 \advance\pmhours by -12
                 \the\pmhours:\minute\ PM
                 \fi
            \fi
      \fi
\fi
}

\def\monthname{\relax\ifcase\month 0/\or January\or February\or
   March\or April\or May\or June\or July\or August\or September\or
   October\or November\or December\else\number\month/\fi}

\def\bold#1{\setbox0=\hbox{$#1$}%
     \kern-.025em\copy0\kern-\wd0
     \kern.05em\copy0\kern-\wd0
     \kern-.025em\raise.0433em\box0 }

\def\gappeq{\mathrel{\rlap {\raise.5ex\hbox{$>$}}
{\lower.5ex\hbox{$\sim$}}}}

\def\lappeq{\mathrel{\rlap{\raise.5ex\hbox{$<$}}
{\lower.5ex\hbox{$\sim$}}}}

\def\ga{\mathrel{\raise.3ex\hbox{$>$\kern-.75em\lower1ex\hbox{$\sim$}}}}
\def\la{\mathrel{\raise.3ex\hbox{$<$\kern-.75em\lower1ex\hbox{$\sim$}}}}
\def\gev{{\rm \, Ge\kern-0.125em V}}
\def\tev{{\rm \, Te\kern-0.125em V}}
\def\beq{\begin{equation}}
\def\eeq{\end{equation}}

\def\m12{m_{1\!/2}}

\begin{document}
\begin{titlepage}
\pagestyle{empty}
\baselineskip=21pt
\rightline{hep-ex/0011086}
\rightline{CERN--TH/2000-307}
%\rightline{ACT-14-00, CTP-TAMU-31/00}
%\rightline{MADPH-00-1166}
%\rightline{UMN--TH--1925/00, TPI--MINN--00/49}
\vskip 1in
\begin{center}
{\large{\bf
The 115 GeV Higgs Odyssey}}
\end{center}
\begin{center}
\vskip 0.5in
%{
{\bf John Ellis}
\vskip 0.4in
{\it Theoretical Physics Division, CERN, Geneva, Switzerland}\\
\vskip 0.4in

{\bf Abstract}
\end{center}
{On his way home from Troy, Odysseus had arrived within reach of Ithaca
when a great storm blew up.  He was swept away, and only several years
later was he able to return to reclaim his rights from the rapacious
suitors, with the aid of his son Telemachus. Some wonder whether this epic is
repeating itself, if the Higgs weighs 115~GeV. If so, 
are CMS and ATLAS cast in the role of Telemachus? 
In this paper, I first discuss how close to Ithaca LEP may have been, 
the fact that a 115~GeV Higgs boson would disfavour technicolour, its
potential
implications for supersymmetry, and finally the prospects for completing
the Higgs Odyssey.}

\vskip 0.4in
\begin{center}
{\it Invited contribution to the CMS Bulletin of December 2000}
\end{center}
\vskip 0.4in
\baselineskip=18pt \noindent
%%%%%%%%%%%%%%%%%%%%%%%%%%%%%%%%%%%%%%%%%%%%%%%%%%%%%%%%%%%%%%%%%%%%%

%%%%%%%%%%%%%%%%%%%%%%%%%%%%%%%%%%%%%%%%%%%%%%%%%%%%%%%%%%%%%%%%%%%%%\vskip 0.15in
\leftline{CERN--TH/2000-307}
\leftline{November 27th, 2000}
\end{titlepage}
\baselineskip=18pt
%%%%%%%%%%%%%%%%%%%%%%%%%%%%%%%%%%%%%%%%%%%%%%%%%%%%%%%%%%%%%%%%%%%%%

\section{How far from Ithaca?}

\begin{figure}[htbp]
\begin{center}
%\hglue1.5cm
%\hskip -2.75cm
\mbox{\epsfig{file=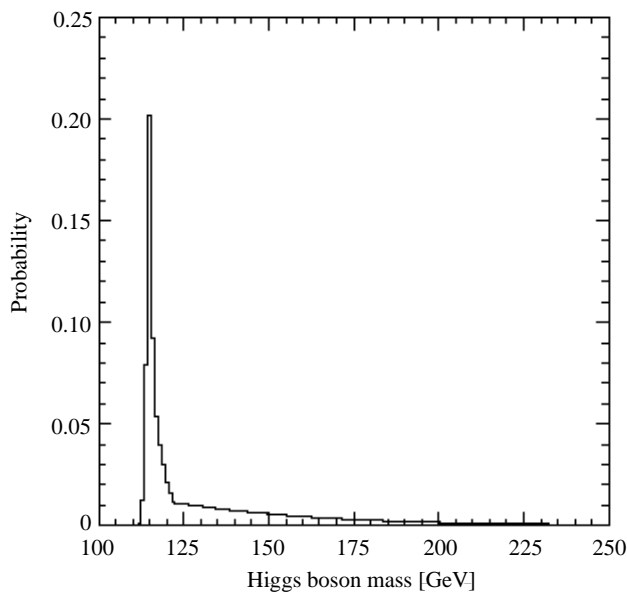,height=8cm}}
\end{center}
%\vskip-7.0cm
\caption[.]{\label{fig:Higgsprob}\it
Probability distribution for the Higgs mass, obtained by
combining the LEP lower limit~\cite{PeterIK} with a precision electroweak
fit~\cite{Erler}.
}
\end{figure}

For several years now, the precision electroweak data have been
suggesting~\cite{LEPEWWG}
that the Higgs boson is very close to the present experimental limit of
113.5~GeV~\cite{PeterIK}.  Fig.~\ref{fig:Higgsprob} shows a recent
combination of the likelihood information from one precision electroweak
fit with the lower limit from direct searches at LEP~\cite{Erler}.  The
central value of $m_H$ indicated by the precision electroweak data
increases by about 30~GeV if the new BES data on $\sigma$ ($e^+e^-
\rightarrow$ hadrons) are used to re-evaluate
$\alpha_{em}(m_Z)$~\cite{LEPEWWG}, but the message remains clear:  Ithaca
cannot be far away.

As is well known, in the minimal supersymmetric extension of the Standard
Model (MSSM), the Higgs mass may be calculated, is unlikely to exceed
130~GeV~\cite{EFZ}, and may well be considerably lighter, depending on the
ratio of Higgs v.e.v.'s and on the mass of the stop squark.  Later we use
this linkage to estimate the sparticle spectrum.  For the moment, we just
note that this is an independent argument for thinking that LEP
might have arrived within sight of Ithaca.

\begin{figure}[htbp]
%\begin{center}
\hglue3.5cm
%\hskip -1cm
\mbox{\epsfig{file=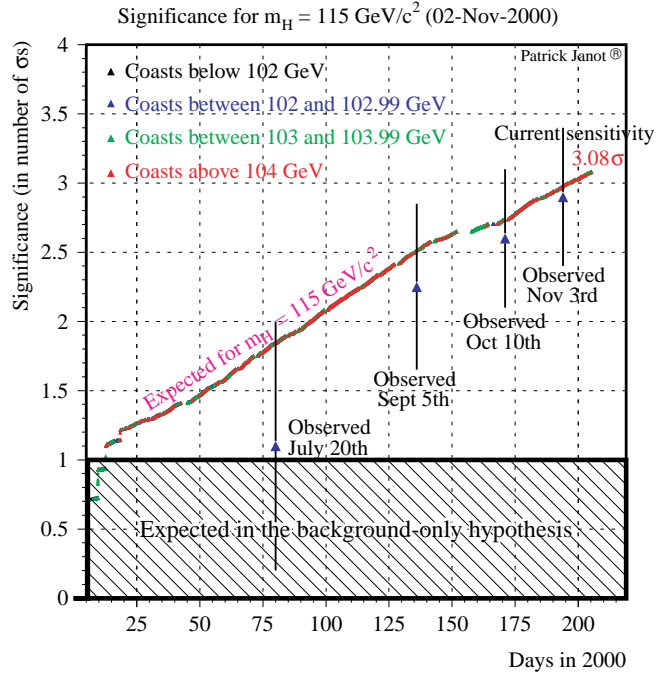,height=9cm}}
%\end{center}
%\vskip-2.0cm
\caption[.]{\label{fig:Janot1}\it 
Expected LEP~2 sensitivity to a Higgs weighing 115~GeV,
as a function of the number of days running at high energy in 2000. Also
shown as
triangles with error bars are the magnitudes of the observed `signals' on
Sept.~5th, Oct.~10th and Nov. 3rd.}
\end{figure}

Much excitement has been generated by the `signal' for a Higgs boson
with $m_H = 115.0^{+1.3}_{-0.7}~{\rm GeV}$ (90\% confidence range),
produced in association with a $Z^0$~\cite{EGN}, reported by the
LEP collaborations~\cite{LEP} and the LEP Working Group on Higgs boson
searches this Autumn. The overall significance on
Nov.~3rd~\cite{PeterIK} was higher than on Oct.~10th~\cite{LEPFest}, which
was in turn higher than on Sept.~5th~\cite{LEPC}:  $2.2 \sigma \rightarrow
2.5 \sigma \rightarrow 2.9 \sigma$.  Moreover, the overall significances
on all three dates were in (improving) agreement with the estimated
sensitivity for $m_H \sim 115$~GeV: see
Fig.~\ref{fig:Janot1}~\cite{Janot}. The probability that the data sample
be compatible with background only was found by the LEP Higgs working
group to be 1.2\% on Sept. 5th, 0.6\% on Oct. 10th, and 0.4\% on Nov. 3rd,
respectively.

The overall significance of the LEP~2
Higgs `signal' did not decrease as would have been expected if there was
only background, but instead grew just as would be
expected if there was also a real Higgs weighing 115~GeV~\cite{PeterIK}.

\begin{figure}
\begin{center}
\mbox{\epsfig{file=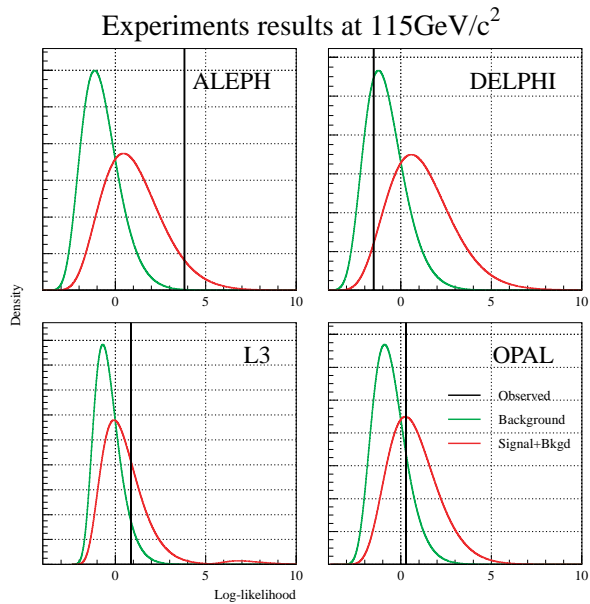,height=8cm}}
\end{center}
%\vskip -2cm
\caption[.]{\label{fig:fourexpts}\it 
The distributions of log-likelihood expected in the background
and signal + background hypotheses in the four LEP experiments, compared
with their observations (vertical lines)~\cite{PeterIK,Janot}.
}
\end{figure}

The growth in significance between September 5th and November 3 was
mainly a
result of the accumulation of interesting candidates by L3 and OPAL. The
DELPHI `signal' reported on September 5th weakened after re-analysis, and
the ALEPH `signal' did not grow. The present situation is that ALEPH
still has the largest `signal', the next largest is in L3, then OPAL, and
DELPHI is more compatible with pure background, as seen in
Fig.~\ref{fig:fourexpts}. The distribution of log-likelihood across the
four experiments is quite consistent with common sampling of the same
`signal'. 

\begin{figure}
\begin{center}
%\hglue3.5cm
%\hskip-6cm
\mbox{\epsfig{file=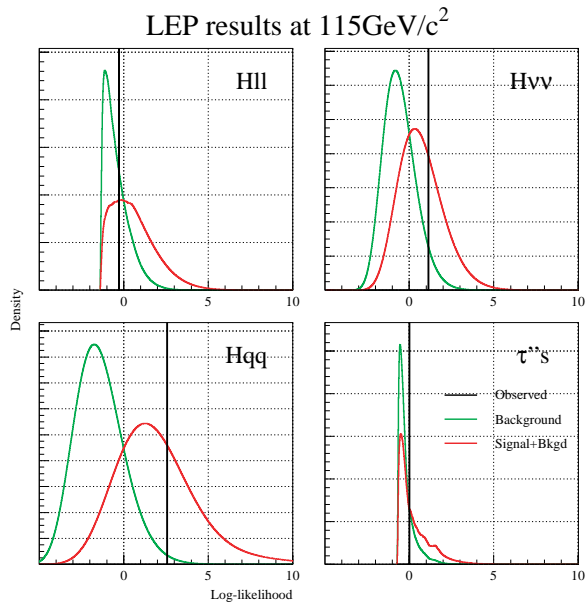,height=8cm}}
\end{center}
\caption[.]{\label{fig:fourchannels}\it 
The distributions of log-likelihood expected in the background
and signal + background hypotheses in the four possible Higgs detection
channels, compared
with their observations (vertical lines)~\cite{PeterIK,Janot}.
}
\end{figure}

Moreover, the `signal' spread from the original ${\bar b} b {\bar q} q$
channel to the ${\bar b} b {\bar \nu} \nu$ channel, where it is now almost
as significant, as seen in Fig.~\ref{fig:fourchannels}.  This metastasis
is largely due to an interesting $\bar b b \bar\nu \nu$ event from L3.
They see a pair of $\bar b b$
jets with relatively low net $p_T$ that emerge almost back to back.  This
would be surprising if their invariant mass were much less than 114~GeV,
but is just what one might expect if $H \rightarrow \bar b b$ with $m_H >
114$~GeV~\cite{L3}, as confirmed by an ALEPH Monte Carlo
simulation~\cite{Wu}. 

Thus the LEP Higgs `signal' did all that it could with the increase in the
statistics analyzed on Nov.~3rd - increased in significance, spread to
other detectors and to another channel. It would have been nice to have a
`gold-plated' event, e.g., in the ${\bar b b} \ell^+ \ell^-$ channel, but
the truth is that no channels are background-free at LEP~\footnote{The
same is even more true at the Tevatron and the LHC.}.

\section{Meanwhile on Mount Olympus}

What would be the significance of a Higgs weighing 115~GeV~\cite{EGNO}?
It would not
just be the crowning confirmation of the Standard Model, but would also be
evidence for new physics beyond it, at a relatively low scale, potentially
accessible to the LHC.  The reason for this is the shape of the effective
Higgs potential, determined by the quartic Higgs coupling $\lambda_H$. 
This is subject to renormalization by the top-quark Yukawa coupling
$\lambda_t$, as well as by the quartic Higgs coupling $\lambda_H$ itself. 
With $m_t \sim 175$~GeV and $m_H \sim 115$~GeV, the renormalization by
$\lambda_t$ is stronger.  Moreover, it tends to decrease $\lambda_H$,
eventually turning it negative at a scale $\lappeq 10^6$~GeV~\cite{AI}.
This
causes the effective Higgs potential to become unbounded below, implying
that our present electroweak vacuum is unstable - {\it unless some new
physics
is introduced at an energy below $10^6$~GeV.}

Could this new physics be a new non-perturbative set of strong
interactions, as in
technicolour or topcolour models? These generally predict large effective
scalar masses, e.g., about 1~TeV in the technicolour case~\cite{TC}.  In
order to
generate fermion masses, one needs to extend technicolour, and such models
predict additional pseudoscalar bosons weighing $\sim 100$~GeV. However,
these would not be produced at LEP in association with the
$Z$~\cite{hole}. 
Therefore, {\it technicolour has no obvious candidate for a 115~GeV
`Higgs'},
and the same seems to be true of other strongly-interacting models of
electroweak symmetry breaking~\cite{EGNO}.

On the other hand, such {\it a light Higgs boson cries out for
supersymmetry},
and vice versa. In any perturbative framework, one can argue that the new
low-energy physics should be {\it bosonic}, so that it may help
$\lambda_H$
counterbalance the destabilizing effects of $\lambda_t$. In 
MSSM, the task of stabilization is
undertaken by the stop squarks.  

\begin{figure}[htbp]
\begin{center}
%\vskip -1.0in
\mbox{%\hskip -.7in 
\epsfig{file=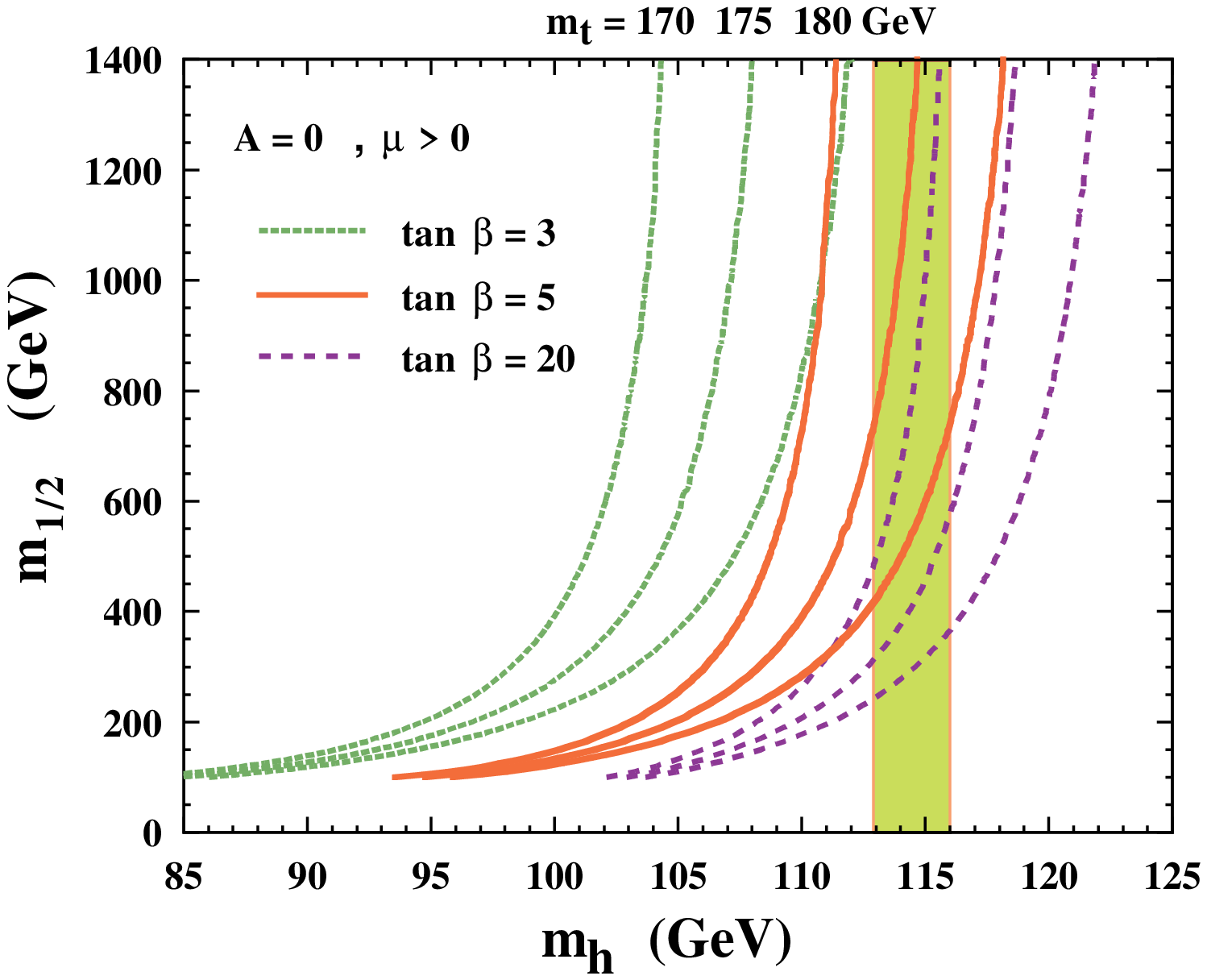,height=7cm}}
%\vskip -2.1in
\mbox{
%\hskip -.7in 
\epsfig{file=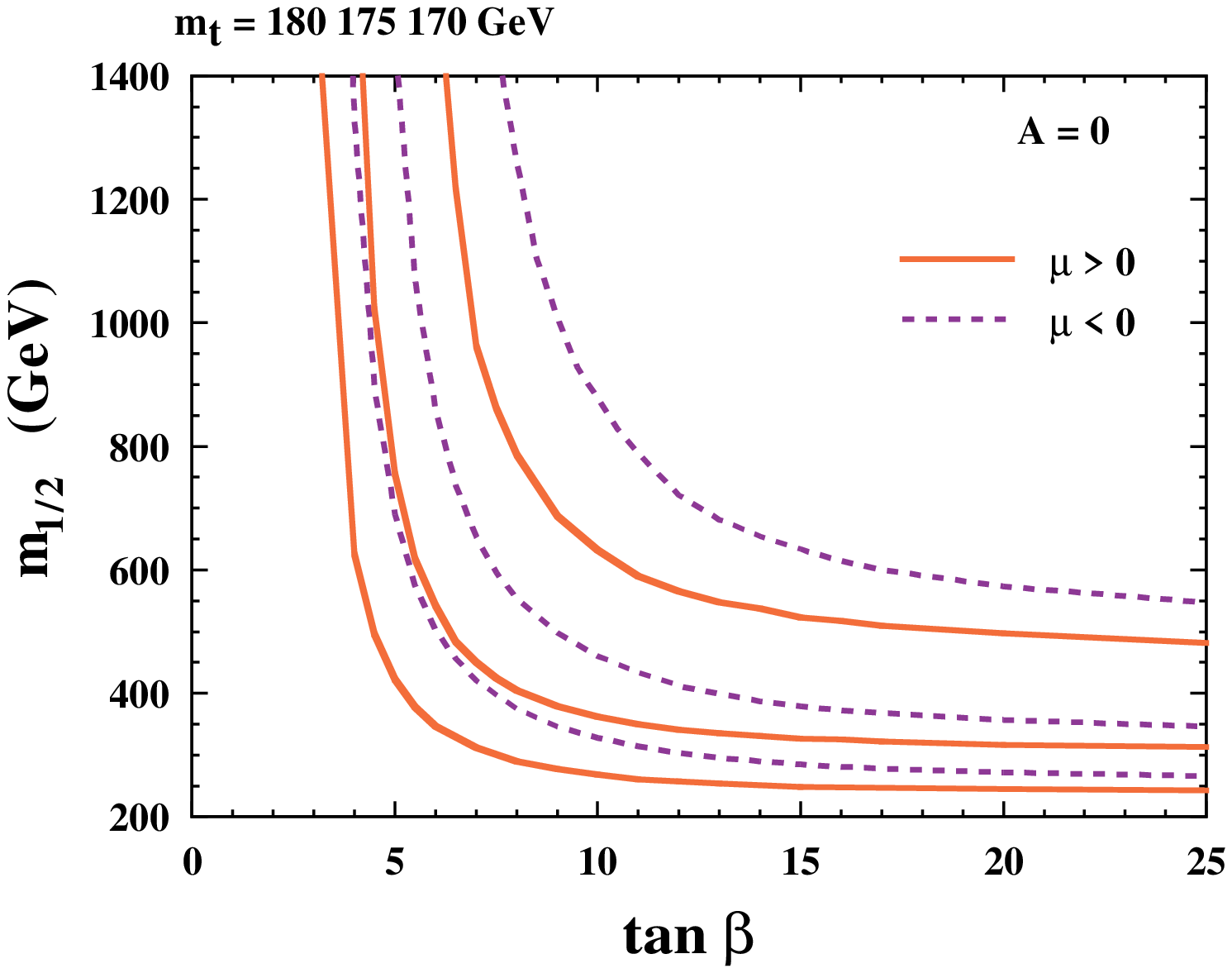,height=7cm}}
\end{center}
%\vskip -1.3in
\caption[.]{\label{fig:EGNO}\it (a) The value of the sparticle
mass scale $m_{1/2}$
required in the minimal supersymmetric
extension of the Standard Model, assuming universal input
sparticle masses, to obtain $m_H \sim 115$~GeV for different
values of $\tan\beta$ and $m_t$, and (b) the corresponding
lower bound on $m_{1/2}$~\cite{EGNO}.} 
\end{figure}

As already mentioned, in the MSSM the
lightest neutral Higgs boson is predicted to weigh $\lappeq 130$~GeV. 
Since its mass is sensitive, via radiative corrections, to sparticle
masses, one can try to use the `measurement' $m_H = 115$~GeV to guess how
heavy squarks and other sparticles might be.  To do this requires some
assumption on the nature of the sparticle spectrum.  For example, if all
the spin-0 sparticles are assumed to be degenerate at some high (GUT) 
energy scale with a mass $m_0$ and likewise for the spin-1/2 gauginos with
a common mass $m_{1/2}$, we found that $m_H$ is most sensitive to
$m_{1/2}$~\cite{EGNO}. Indeed, as seen in Fig.~\ref{fig:EGNO}, we found
the lower limit $m_{1/2} \gappeq 240$~GeV if $m_H
\gappeq 113$~GeV and $m_t \leq 180$~GeV.  The gluino and squark masses
would then be 2 or 3 times heavier:  $m_{\tilde g} \gappeq 600$~GeV and
$m_{\tilde q} \gappeq 700$~GeV, beyond the reach of the
Tevatron~\footnote{These lower bounds remain valid even if the LEP Higgs
`signal' eventually turns out to be a chimera. Other groups have repeated
this analysis using alternative assumptions, obtaining analogous
results~\cite{others}.}. However, masses up to a factor three above these
lower limits
are within reach of the LHC, which should be able to cover all of the
$(m_0, m_{1/2})$ parameter region where the lightest supersymmetric
particle is likely to constitute the cold dark matter posited by
astrophysicists and cosmologists~\cite{EGNO}.

\section{Return to Ithaca}

\begin{figure}
\begin{center}
%\vskip3.0cm
\hglue1.0cm
\mbox{\epsfig{file=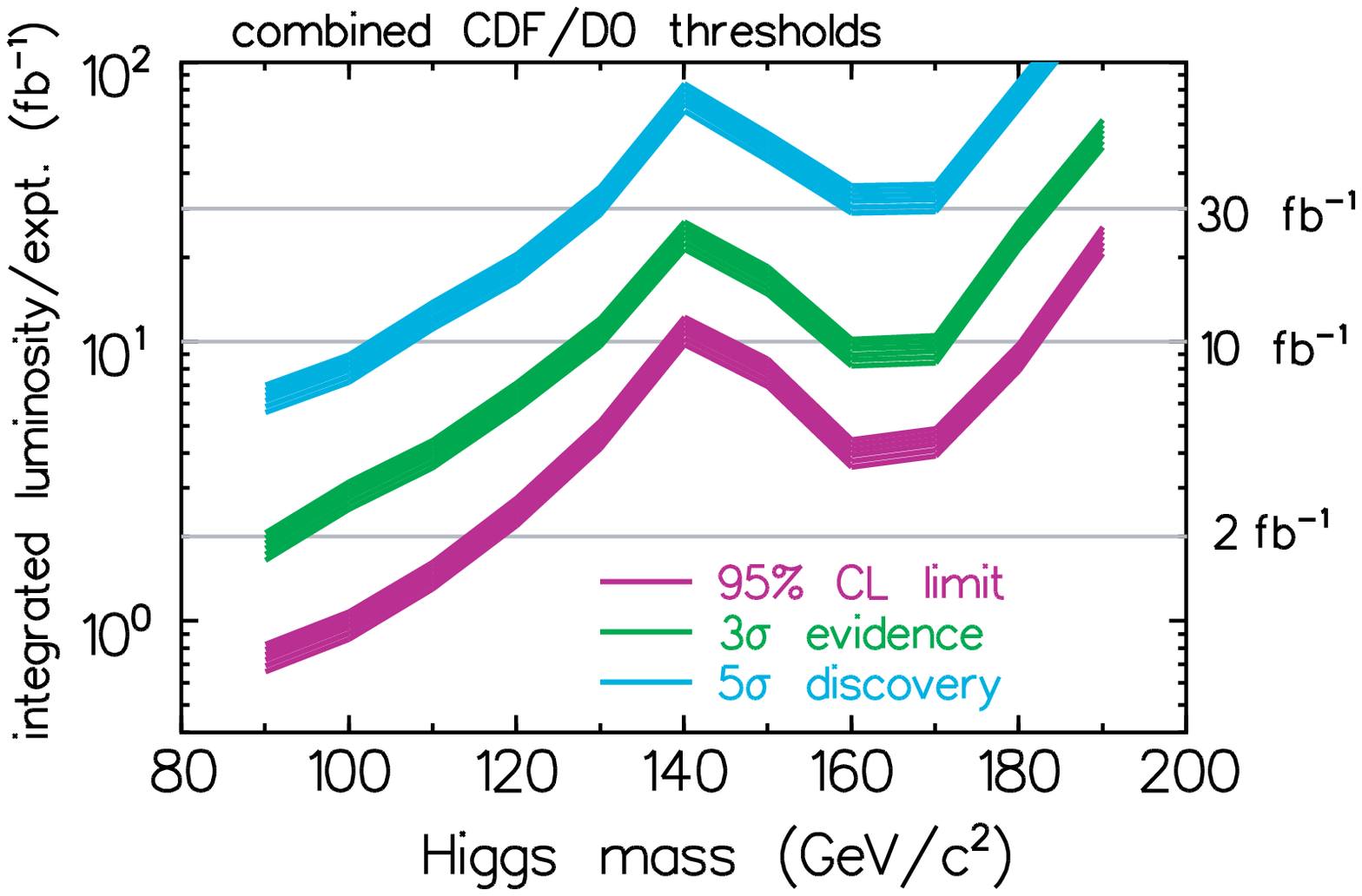,height=6.5cm}}
\end{center}
%\vskip-4cm
\caption[.]{\label{fig:Tevatron1}\it
The sensitivity of the FNAL Tevatron experiments to a light
Higgs, as a function of its mass~\cite{THiggs}.}
\end{figure}

We now turn to the jealous suitors and Telemachus.  What is the
sensitivity of the Tevatron experiments to a Higgs weighing 115~GeV? As
seen in Fig.~\ref{fig:Tevatron1}, in order to attain 3 (5)~$\sigma$, it
is estimated that they would need 5 (15)~fb$^{-1}$ \cite{THiggs}.  As for
the prospective Tevatron luminosity, at the moment, 2~fb$^{-1}$ is
`promised' by 2003. However, a roadmap for reaching 15~fb$^{-1}$ by 2007
has been proposed. If this is achieved, the Tevatron may have a chance if
the Higgs weighs 115~GeV, but does not seem likely to detect any
heavier Higgs boson.

\begin{figure}[htbp]
\begin{center}
%\hglue3.0cm
\vskip-2cm
\mbox{\epsfig{file=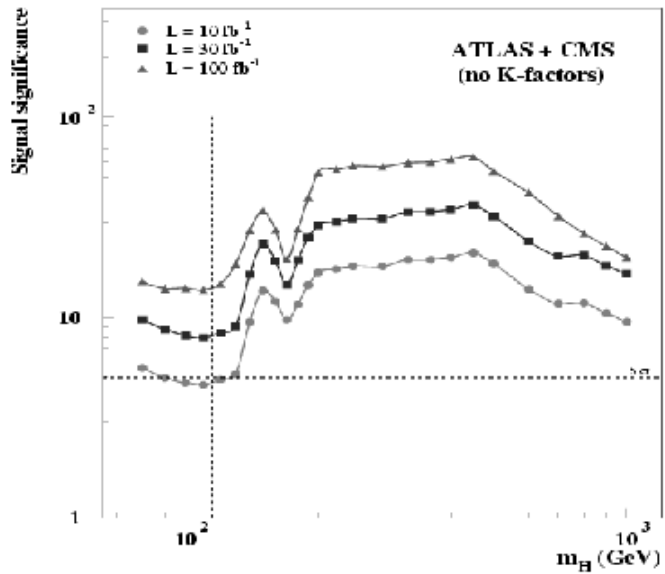,height=8cm}}
\end{center}
\vskip 2cm
\caption[.]{\label{fig:LHCsignificance}\it
The sensitivity of the LHC experiments to a 
Higgs, as a function of its mass, for different accumulated
luminosities~\cite{Gianotti}.}
\end{figure}

What of Telemachus?  According to CMS and ATLAS studies, as seen in
Fig.~\ref{fig:LHCsignificance}~\cite{Gianotti}, the minimum luminosity
required to start seeing a 115~GeV Higgs at 5 $\sigma$ is $\sim
10$~fb$^{-1}$, which may be achieved after two years of LHC running. Since
at most a few weeks of very low luminosity collisions can be envisaged in
2005, and only 1 or 2~fb$^{-1}$ is anticipated in 2006, this presumably
means that the LHC could hope to discover a 115~GeV
Higgs boson after the 2007 run.

\vskip 1cm

There is, however, an important proviso. The LEP production mechanism,
$e^+ e^- \rightarrow Z + H$~\cite{EGN}, measures a different coupling -
$ZZH$ - from those to which the LHC is sensitive - 
${\bar t} t H$ and $\gamma \gamma H$, with the latter being quite
model-dependent. The
$\gamma \gamma H$ coupling is controlled by loop diagrams sensitive to
virtual particles and the $\bar t t H$ coupling is sensitive to the ratio
of v.e.v.'s and Higgs mixing in the MSSM. Therefore the information
obtained at the Tevatron and the LHC will be complementary to that
obtained by LEP, and both sets of information will be helpful in
determining whether the candidate Higgs boson has all the expected
couplings.

The long-term plans for high-energy physics at all major laboratories
around the world (NLC, JLC, TESLA, Muon Collider) depend very much whether
or not there is a light Higgs. All the indications from LEP precision data
are that it must weigh $\la 200$~GeV~\cite{LEPEWWG}. A Higgs in the hand
would be worth two in the bush to the NLC, JLC, TESLA and Muon Collider
communities, when they approach their funding agencies. For this they may
have to wait until 2007. Until then, they and the rest of the particle
physics community may be left in suspense while the Higgs Odyssey
continues across the wine-dark seas of LHC construction.

In the mean time - even if we are cast in subordinate roles analagous to
the swineherd Eumaeus, old Laertes, faithful Penelope, the observant nurse
Eurycleia, or the dying dog Argos - we all support CMS and ATLAS in their
definitive search for the Higgs boson.

\end{document}